\documentstyle{mn}
\title{Efficiency of cosmic ray reflections from an ultrarelativistic 
shock wave}

\author[J. Bednarz \& M. Ostrowski]
       {J. Bednarz, M. Ostrowski \\
Obserwatorium Astronomiczne, Uniwersytet Jagiello\'nski, ul. Orla 171,
30-244 Krak\'ow, Poland}

\begin{document}

\maketitle

\label{firstpage}

\begin{abstract}
The process of cosmic ray acceleration up to energies in excess of
$10^{20}$ eV at relativistic shock waves with large Lorentz factors,
$\Gamma \gg 1$ requires $\sim \Gamma^2$ particle energy gains at single
reflections from the shock (cf. Gallant \& Achterberg 1999). In the
present comment, applying numerical simulations we address an efficiency
problem arising for such models. The actual efficiency of the
acceleration process is expected to be substantially lower than the
estimates of previous authors.

\end{abstract}


\section{Introduction}

In attempt to substitute a single question mark for the previous two,
some authors try to identify the process accelerating particles to ultra
high energies (`UHE', energy $E > 10^{18}$ eV) with ultrarelativistic
shock waves considered to be sources of $\gamma$-ray bursts (cf. Waxman
1995a,b, Vietri 1995, Milgrom \& Usov 1995, Gallant \& Achterberg 1999,
$\equiv$ GA99). In the proposed models, reaching cosmic ray energies in
excess of $10^{20}$ eV requires that particles are reflected from the
shock wave characterized with a large Lorentz gamma factor, $\Gamma$, to
enable a relative energy gain -- in a single reflection -- comparable to
$\Gamma^2$ (cf. GA99). On the other hand, basing on our experience with
numerical modeling, we suggested (Bednarz \& Ostrowski 1998, Ostrowski
1999) that such processes cannot actually work due to low efficiency of
particle reflections.

In the present note we elaborate this problem in detail with the use of
numerical modeling of particle interactions with the shock. We show that
`$\Gamma^2$' reflections can occur with a non-negligible rate only if a
substantial amount of turbulence is present downstream of the shock.
However, even in such conditions the number of accelerated particles is
a small fraction of all cosmic rays hitting the shock.

\section{Simulations}

As discussed by Bednarz \& Ostrowski (1998), and in detail by GA99,
particles accelerated in multiple interactions with an ultrarelativistic
shock wave gain on average in a single `loop' --
upstream-downstream-upstream -- the amount of energy comparable to the
original energy, $<\Delta E> \approx E$. It is due to extreme particle
anisotropy occurring in large $\Gamma$ shocks. Particles hitting the
shock wave the first time, with their isotropic upstream distribution,
can receive higher energy gains.  In this case an individual reflection
from the shock may increase particle energies on a factor of $\sim
\Gamma^2$. However the effectiveness of such acceleration depends on how
many of particles from the original upstream population can be
reflected. To verify it we use Monte Carlo  simulations similar to the
applied earlier to derivation of the accelerated particle spectra
(Bednarz \& Ostrowski 1998, for details see also Bednarz \& Ostrowski
1996). The code reproduces perturbations of particle trajectories due to
MHD turbulence by applying discrete scatterings of particle direction
within a narrow cone along its momentum vector. A procedure uses a
hybrid approach involving very small scattering angles close to the
shock and larger angles further away from it. Between successive
scatterings particle trajectories are derived in the uniform background
magnetic field. The respective scaling of the time between the
successive scattering acts close and far from the shock mimics the same
turbulence amplitude everywhere. The scattering amplitude was selected
in a way to reproduce a pitch angle diffusion process for particle
momentum. It requires the angular scattering amplitude of particle
momentum vector to be much smaller than the particle anisotropy. All
computations were done in a respective local plasma rest frame and the
Lorentz transformation was applied to every particle crossing of the
shock.

Particles with initial momenta $p_0$ taken as a momentum unit,
$p_{0}=1$, were injected  at the distance of $2r_{g}$ ($r_{g}$ -
particle gyroradius) upstream of the shock front. For all such particles
we derived their trajectories until they crossed the shock upstream, or
were caught in the downstream plasma, reaching a distance of $4 r_g$
downstream of the shock. For each single particle interaction with the
shock the particle momentum vector was recorded, so we were able to
consider angular and energy distributions of such particles. We
considered shocks with Lorentz factors $\Gamma= 10$, $160$ and $320$.
For each shock we discussed the acceleration processes in conditions
with the magnetic field inclination $\psi= 0^\circ$, $10^\circ$,
$70^\circ$ and with 16 values for the turbulence amplitude measured by
the ratio $\tau$ of the cross-field diffusion coefficient,
$\kappa_\perp$, to the parallel diffusion coefficient, $\kappa_\|$. The
applied values of $\tau$ were taken from the range of ($3.2\cdot
10^{-6}$, $0.95$), approximately uniformly distributed in $\log \tau$~.
In each simulation run we derived trajectories of $5\cdot 10^{4}$
particles with initial momenta isotropically distributed in the upstream
rest frame.

\section{Efficiency of `$\Gamma^2$' reflections}

In the downstream plasma rest frame the ultrarelativistic shock moves
with velocity $c/3$. This velocity is comparable to the particle
velocity $c$. Therefore, from all particles crossing the shock
downstream only the ones with particular momentum orientations will
interact with the shock again, the remaining particles will be caught in
the downstream plasma flow and advected far from the shock front. In the
simulations we considered this process quantitatively. However, let us
first present a simple illustration.

Large compression ratios occurring in ultrarelativistic shocks, as
measured between the upstream and downstream plasma rest frames, lead
for nearly all oblique upstream magnetic field configurations to the
quasi-perpendicular configurations downstream of the shock. Thus, let us
consider for this illustrative example a shock with a non-perturbed
perpendicular downstream magnetic field distribution. Particle crossing
the shock downstream with inclination to the magnetic field $\theta$ and
the phase $\phi$ -- both measured in the downstream plasma rest frame,
$\phi = \pi/2$ for particles normal to the shock and directed downstream
-- will be able to cross the shock upstream only if the equation

$$ {c \over 3} \, t = r_{\rm g} \left[ \cos (\phi + \omega_{\rm g} t ) 
- \cos \phi \right] \eqno(3.1)$$

\noindent
has a solution at positive time $t$. Here $r_{\rm g} = { pc \over eB}
\sin \theta$ is the particle gyroradius, $\omega_{\rm g} = {e B \over
p}$ is the gyration frequency, and other symbols have the usual meaning.

\begin{figure}
\vspace*{5.5cm}
\includegraphics{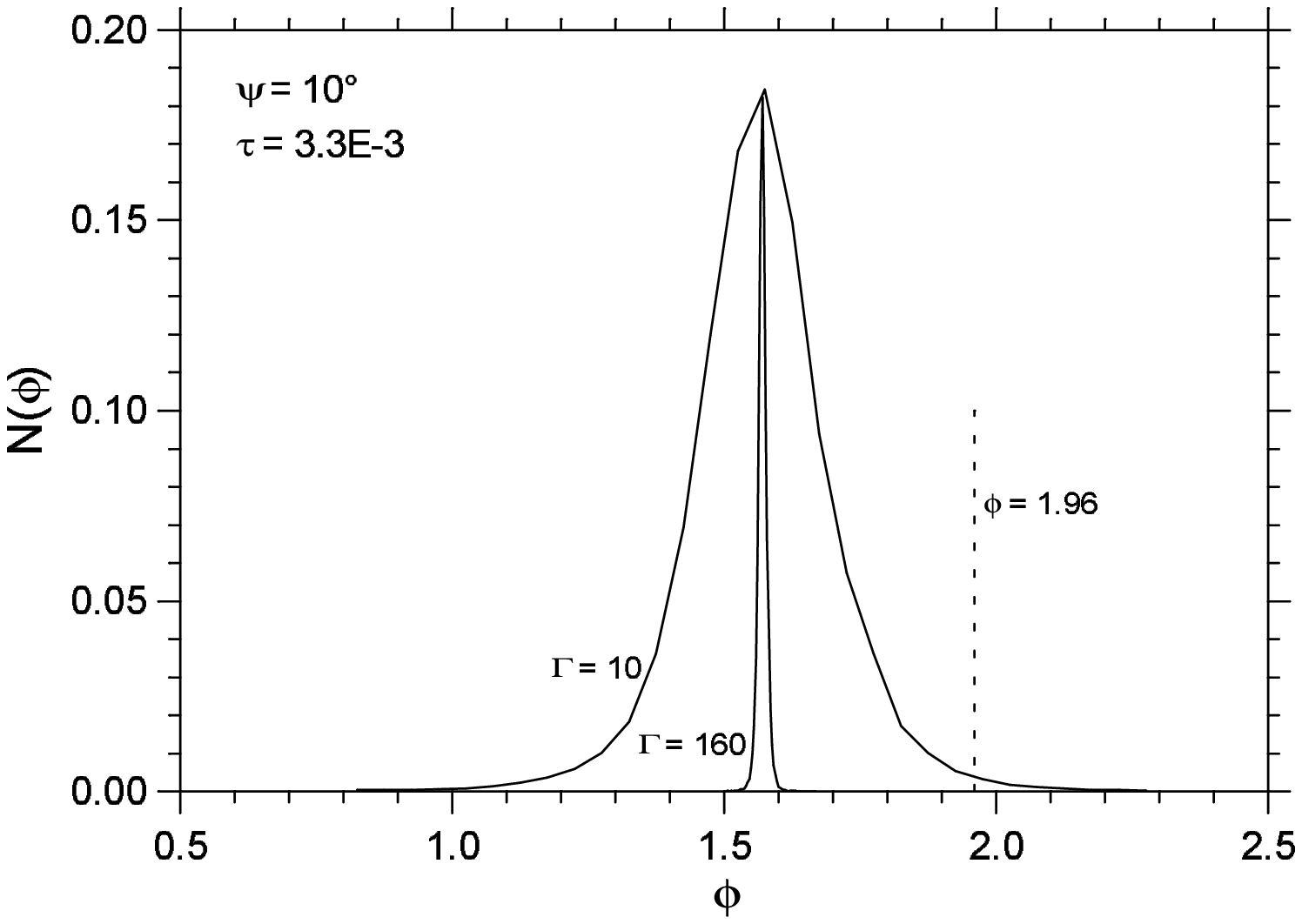}
\caption{A distribution of particle phases for particles crossing 
the shock downstream (as measured in the downstream plasma rest frame), 
if their upstream distribution was isotropic. 
A dashed line delimits a range of particle phases below which particles
are not able to reach the shock again at the perpendicular uniform 
downstream magnetic field. }
\label{fig1}
\end{figure}

An angular range in the space ($\theta$, $\phi$) enabling particles
crossing the shock downstream to reach the shock again can be
characterized for illustration by three values of $\theta$. Particles
with $\sin \theta = 1$ are able to reach the shock again if $\phi
\in$(1.96, 3.48), with $\sin \theta = 0.5$ if $\phi \in$(2.96, 3.87) and
with $\sin \theta = 1/3$ only for $\phi = 4.71$. That means that all
particles with $\phi$ smaller than 1.96 (Fig.~1) are not able to reach
the shock again if fluctuations of the magnetic field downstream of the
shock are not present.

\begin{figure}
\vspace*{5.5cm}
\includegraphics{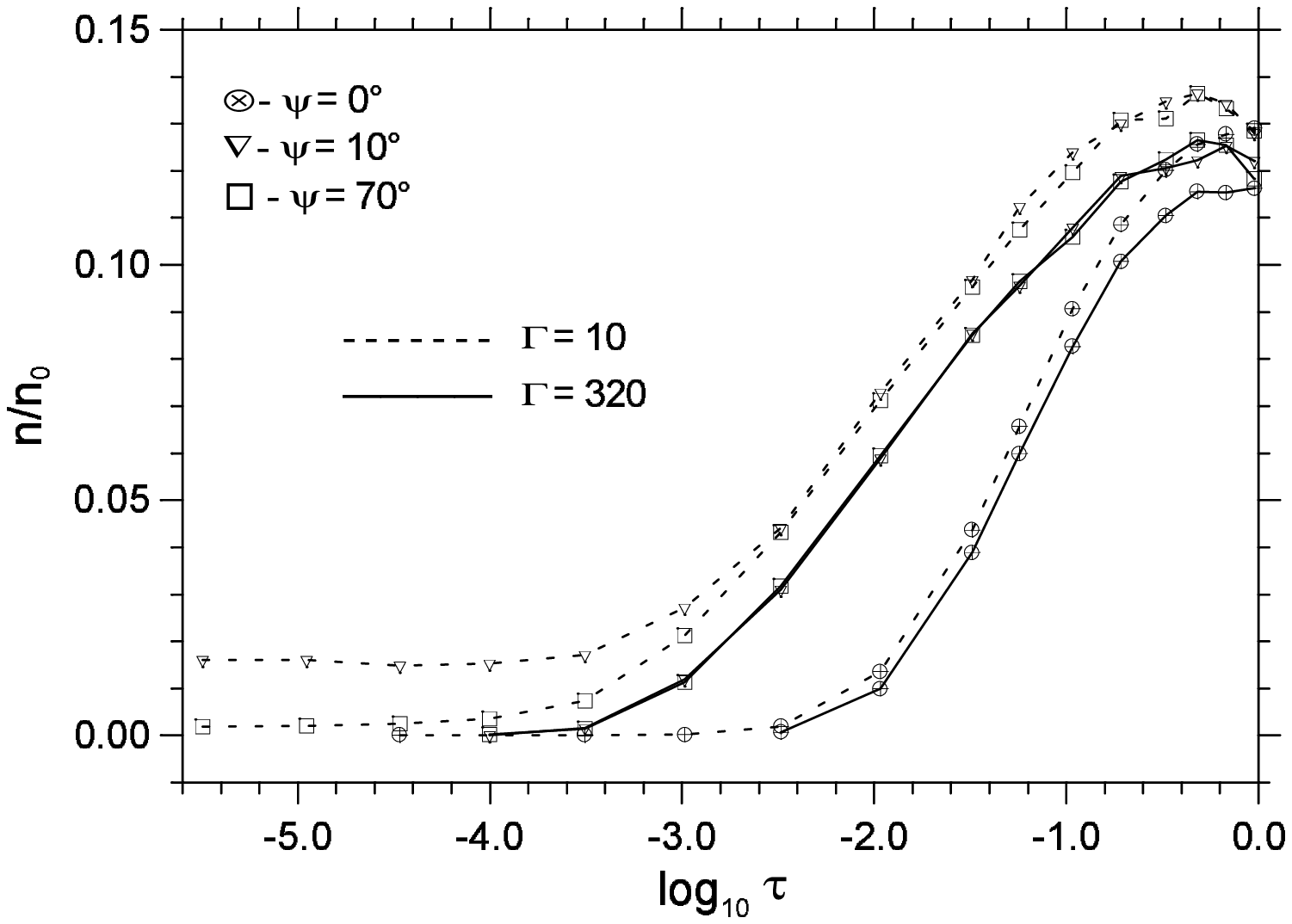}
\caption{
A ratio of the number of reflected particles, $n$, to all particles 
crossing the shock downstream, $n_0$, as a function
of the magnetic field fluctuations amplitude, $\tau$.}
\label{fig2}
\end{figure}

\begin{figure}
\vspace*{5.5cm}
\includegraphics{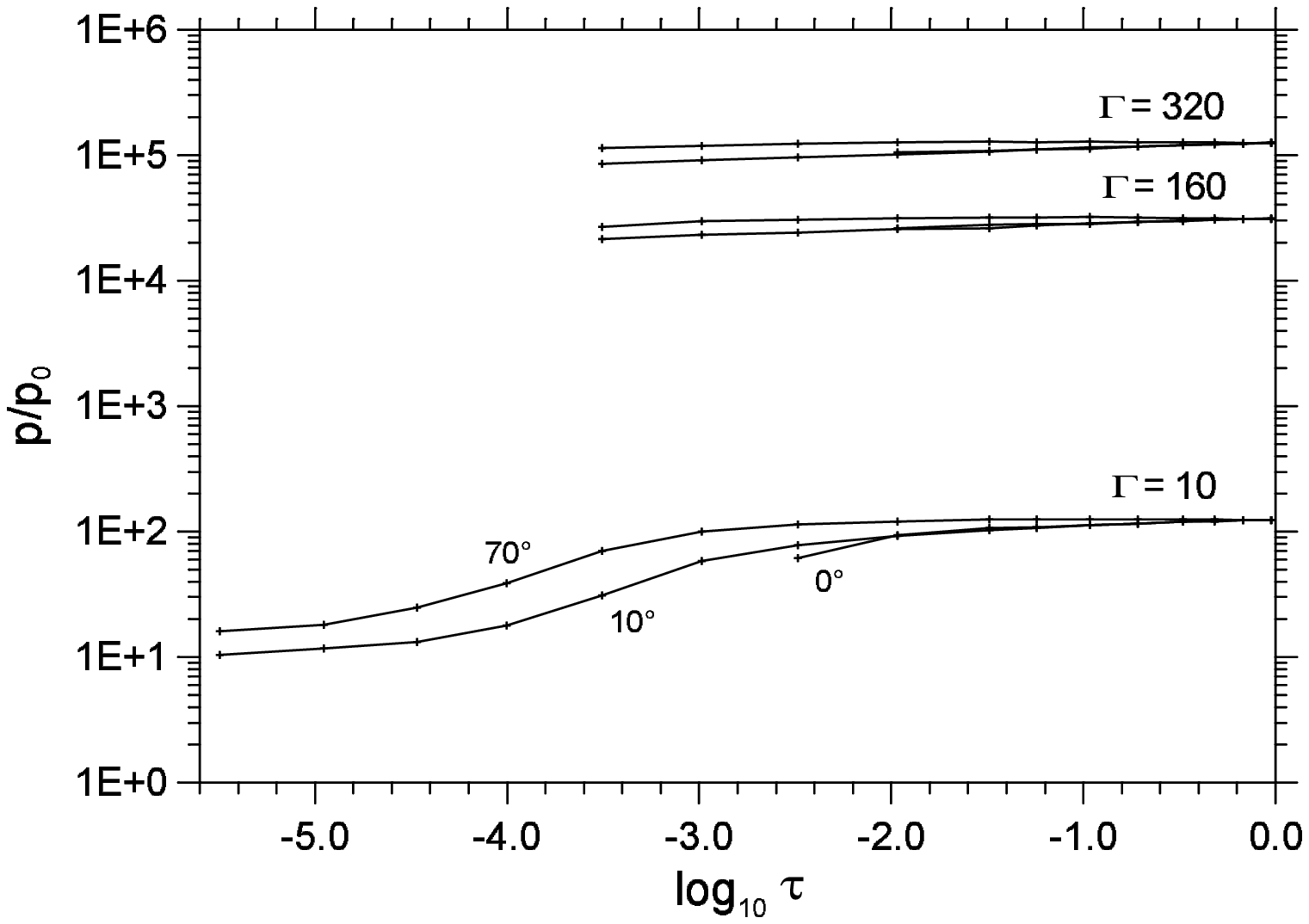}
\caption{
Momentum gains of reflected particles, $p/p_0$, as a function of the
magnetic field fluctuations amplitude, $\tau$. For large magnetic field
fluctuations the momentum gain approaches $\approx 1.2 \Gamma^{2}$
independently of the shock Lorentz factor.}
\label{fig3}
\end{figure}

For perturbed magnetic fields some downstream trajectories starting in
the ($\theta$, $\phi$) plane outside the reflection range can be
scattered toward the shock to cross it upstream. We prove it by
simulations presented in Fig.~2. One may observe that increasing the
perturbation amplitude leads to increased number of reflected particles,
reaching $\approx 13$\% in the limit of $\tau = 1$. For large magnetic
field fluctuations the mean relative energy gains of reflected particles
are close to $1.2 \Gamma^2$ for the shock Lorentz factors considered.
One may note that for small $\psi$ and $\Gamma$ the energy gain
increases with growing $\tau$. The points resulting from simulations for
the smallest values of $\tau$ were not included into Fig.~3 because of
small number of reflected particles (cf. Fig.~2).

\section{Summary}

We have shown that efficiency of `$\Gamma^2$' reflections in
ultrarelativistic shock waves strongly depends on fluctuations of
magnetic field downstream of the shock. In the most favorable conditions
with high amplitude turbulence downstream the shock the reflection
efficiency is a factor of 10 or more smaller than the values assumed by
other authors. Moreover, due to the magnetic field compression at the
shock, we do not expect the required large values of
$\kappa_\perp/\kappa_\parallel$ to occur behind the shock (cf. a
different approach of Medvedev \& Loeb 1999). Therefore, with the actual
efficiency of 1 - 10 \% there is an additional difficulty for models
postulating UHE particle acceleration at GRB shocks (cf. GA99). Let us
note, however, that the mean downstream trajectory of the reflected
particle involves only a fraction of its gyroperiod. Thus the presence
of compressive long waves in this region, leading to non-random
trajectory perturbations could modify our estimates.

\section*{Acknowledgements}

The presented computations were partly done on the HP Exemplar S2000 in
ACK `CYFRONET' in Krak\'ow. We acknowledge support from the {\it Komitet
Bada\'n Naukowych\/} through the grant PB 179/P03/96/11~.

\label{lastpage}

\end{document}